\documentclass[]{aa}

\usepackage{color }
\usepackage{natbib}
\usepackage{amsmath}
\usepackage{amssymb}
\usepackage{graphicx}

\newcommand{\bl}[1]{\mbox{\boldmath$ #1 $}}

\title{Accretion bursts in low-metallicity protostellar disks}
\titlerunning{Accretion bursts at sub-solar metallicity}
\author{Eduard I. Vorobyov\inst{1,2,3}, Vardan G. Elbakyan\inst{3}, Kazuyuki Omukai\inst{4}, Takashi Hosokawa\inst{5}, Ryoki Matsukoba\inst{4},  and Manuel Guedel\inst{1}}
\authorrunning{Vorobyov et al.}
   
\institute{ 
    University of Vienna, Department of Astrophysics, T\"urkenschanzstr. 17, Vienna, 1180, Austria \\
    \email{eduard.vorobiev@univie.ac.at} 
    \and 
    Ural Federal University, 51 Lenin Str., 620051 Ekaterinburg, Russia
    \and
     Research Institute of Physics, Southern Federal University, Rostov-on-Don, 344090 Russia
     \and
    Astronomical Institute, Graduate School of Science, Tohoku University, Aoba, Sendai, Miyagi 980-8578, Japan   \and
    Department of Physics, Kyoto University, Sakyo-ku, Kyoto 606-8502, Japan
    }


\date{}

\begin{document}

 \abstract
   {}
   {The early evolution of protostellar disks with metallicities in the $Z=1.0-0.01~Z_\odot$ range was studied with a particular emphasis on the strength of gravitational instability and the nature of protostellar accretion in low-metallicity systems.}
   {Numerical hydrodynamics simulations in the thin-disk limit were employed that feature separate gas and dust temperatures, and disk mass-loading from the infalling parental cloud cores. Models with cloud cores of similar initial mass and rotation pattern, but distinct metallicity were considered to distinguish the effect of metallicity from that of initial conditions.}
   { The early stages of disk evolution in low-metallicity models are characterized by vigorous gravitational instability and fragmentation. Disk instability is sustained by continual mass-loading from the collapsing core. The time period that is covered by this unstable stage is much shorter in the $Z=0.01~Z_\odot$ models as compared to their higher metallicity counterparts thanks to the higher mass infall rates caused by higher gas temperatures (that decouple from lower dust temperatures)  in the inner parts of collapsing cores.
   Protostellar accretion rates are highly variable in the low-metallicity models reflecting a highly dynamical nature of the corresponding protostellar disks. The low-metallicity systems feature short, but energetic episodes of mass accretion caused by infall of inward-migrating gaseous clumps that form via gravitational fragmentation of protostellar disks. These  bursts seem to be more numerous and last longer in the $Z=0.1~Z_\odot$ models in comparison to the $Z=0.01~Z_\odot$ case.
 }
   {Variable protostellar accretion with episodic bursts is not a particular feature of solar metallicity disks. It is also inherent to gravitationally unstable disks with metallicities up to 100 times lower than solar. }

\maketitle

\section{Introduction}
Circumstellar disks are a key element in the theory of star and planet formation. They form during the gravitational collapse of prestellar cloud cores and disperse as the host stars evolve towards the main sequence.
Mass transport in circumstellar disks determines the terminal stellar mass and dust growth in the disk sets the stage for planet formation.  

Among many processes that operate in circumstellar disks, gravitational instability (GI) presents a particular interest in the context of protostellar accretion and planet formation. GI is a dominant mechanism of mass transport in the early protostellar stage of disk evolution \citep{1990LinPringle,2009VorobyovBasu}, thus largely setting the rate of mass accretion on the star. Moreover, the character of mass accretion versus time can also be affected by GI either directly through perturbations caused by global spiral arms \citep{2016Elbakyan} or indirectly by assisting the magnetorotational instability in the disk innermost regions \citep[e.g.,][]{2009ZhuHartmannGammie,2014BaeHartmann}. The role of GI in disk evolution is versatile and it can even affect the chemical composition of circumstellar disks \citep{2015Evans}.

The extreme manifestation of GI - disk gravitational fragmentation - is another important phenomenon that can affect the character of mass accretion on the growing protostar. Inward-migrating clumps formed via disk fragmentation bring a large amount of mass to the disk inner regions, triggering mass accretion bursts when destroyed by tidal torques \citep{2005VorobyovBasu,2015VorobyovBasu,2017MeyerVorobyov}. Disk fragmentation is also considered as a viable and, perhaps, dominant scenario for the formation of giant planets on wide orbits where disk conditions can be favourable for gravitational fragmentation \citep{2003Boss,2003JohnsonGammie,2007MayerLufkin,2009Boley,2013Vorobyov}. When clump migration and  dust settling are taken into consideration, disk fragmentation can also account for the formation of icy and rocky planets at closer orbits \citep{1998Boss,2010Nayakshinb, 2017Nayakshin,2010BoleyHayfield, 2019VorobyovElbakyan}.  It is not clear yet if disk fragmentation can explain the majority of the known close-in giant planets, for which core accretion remains a preferable scenario \citep{1996Pollack,2005Alibert}. 

GI and disk fragmentation have been extensively studied for solar-metallicity disks with respect to mass transport and planet formation. When lower-metallicity environments are considered, the importance of GI in setting the mass accretion rate and the role of disk fragmentation as a planet formation mechanism becomes less clear. Observations of accretion in low-metallcity galactic and extragalactic star-forming regions showed
that accretion rates on the low-metallicity stars are on average higher than in their solar-metallicity counterparts \citep{2012Spezzi,2013DeMarchi}. On the other hand, \citet{2015Kalari} found no difference between the mass accretion rates inferred for the low-metallicity ($1/5Z_\odot$) open cluster Dolidze~25 and solar-metallicity Galactic pre-main-sequence stars in the same mass range. The reason for this difference remains to be understood.

 Numerical hydrodynamics simulations of primordial (Z=0) disks indicate that they are vigorously
GI-unstable and produce many clumps which quickly migrate to the star, triggering accretion and
luminosity outbursts \citep{2013VorobyovDeSouza,2016Sakurai}.
As the disk metallicity increases to $Z=10^{-5}-10^{-3}~Z_\odot$, steady-state models also predict that circumstellar disks should be strongly gravitationally unstable  \citep{2014Tanaka}. However, at subsolar metallicites of $Z\sim 0.1-1.0~Z_\odot$ studies of circumstellar disks produce conflicting results. For instance, \citet{2006Cai} found that GI is stronger in lower-metallicity disks, but disks nevertheless do not fragment. Conversely, \citet{2002Boss} reported that the propensity of circumstellar disks to gravitational fragmentation is largely insensitive to metallicity variations by a factor of 10 on both sides from the solar value, because the disk thermal balance is mainly determined by the radiative input from the central star. Furthermore, \citet{2010MeruBate} also found that reducing metallicity below the solar value promotes disk fragmentation since lower opacities (associated with lower metallicities) allow the disk to cool quicker.  

In this work, we revisit the issue of protostellar accretion and gravitational fragmentation in disks with subsolar metallicities using numerical hydrodynamics simulations in the thin-disk limit. Unlike previous studies, we consider a wider range of metallicities from $Z=10^{-2} Z_\odot$ to $Z=1.0 Z_\odot$. In addition, we employ a sophisticated thermal scheme \citep{2020VorobyovMatsukoba},  which allows decoupling of gas and dust temperatures at low metallicities. Finally, the disk hydrodynamics simulations are linked with a stellar evolution code, which allows us to compute the time- and metallicity-dependent stellar radiative input as the disk evolves with time. The paper is organized as follows. In Sect.~\ref{model} we review the main aspects of disk modeling at metallicities lower than solar. In Sect.~\ref{predisk} we present the initial stages of gravitational collapse for different metallicities. In Sect.~\ref{SolarZ} we discuss the disk evolution in the context of solar metallicity models. Sect.~\ref{LowZ} presents the results of disk modeling and mass accretion rates at low metallicities. Sect.~\ref{summary} summarizes our main results.


\section{Model description}
\label{model}
We use numerical hydrodynamics simulations in the thin-disk limit to model the evolution of protostellar disks with different metallicities. The model was described in detail in \citet{2020VorobyovMatsukoba}. Here, we only review the key aspects of disk modeling at low metallicities. 

The numerical model takes disk self-gravity and turbulent viscosity into account. The turbulent viscosity is described via  the usual Shakura \& Sunyaev parameterization with the $\alpha$-value in the $10^{-4}-10^{-2}$ limits. Unlike many previous studies of disk evolution that make no distinction between gas and dust temperatures, we employ a sophisticated thermal evolution scheme that allows separate gas and dust temperatures and 
considers the following energetic processes:  radiative continuum cooling (or heating) of gas and dust (including collisional transfer of energy between gas and dust), H$_{2}$  and HD line cooling, chemical cooling/heating through considered chemical reactions, metal line cooling, stellar and background irradiation, and viscous heating. Sub-solar metallicities are set by scaling down the gas and dust opacities, dust-to-gas mass ratio, and metal content of the solar-metallicity disk by the corresponding factor.

The pertinent hydrodynamic equations for mass, momentum, and internal energy density in the thin-disk limit can then be written as:
\begin{equation}
\frac{\partial\Sigma}{\partial t}=-\bl{\nabla}_{p}\cdot\left(\Sigma \bl{v}_{p}\right),\label{eq:mass}
\end{equation}

\begin{equation}
\frac{\partial}{\partial t}\left(\Sigma \bl{v}_{p}\right)+\left[\bl{\nabla} \cdot \left(\Sigma 
\bl{v}_{p}\otimes \bl{v}_{p}\right)\right]_{p}=-\bl{\nabla}_{p} P+\Sigma \bl{g}_{p}+\left(\bl{\nabla}\cdot
\bl{\Pi}\right)_{p},\label{eq:momentum}
\end{equation}

\begin{equation}
\frac{\partial e}{\partial t}+\bl{\nabla}_{\mathrm{p}}\cdot\left(e v_{p}\right)=-P\left(\bl{\nabla}_{p} \cdot v_{p}\right) - Q_{\rm tot} + \left(\bl{\nabla} \cdot v\right)_{pp'}: \bl{\Pi}_{pp'},
\label{eq:energy3}
\end{equation}
where subscripts $p$ and $p'$ refer to the planar components $(r,\phi)$
in polar coordinates, $\Sigma$ is the gas mass surface density, $P$ is the vertically integrated
gas pressure calculated via the ideal equation of state as $P=(\gamma-1)e$, $e$ is the internal energy per surface area, $\gamma$ is the ratio of specific heats, $\bl{v}_{p}=v_{\mathrm{r}}\hat{\bl{r}}+v_{\mathrm{\phi}}\hat{\bl{\phi}}$
is the velocity in the disk plane, $\bl{g}_{p}=g_{\mathrm{r}} \hat{\bl{r}} 
+ g_{\mathrm{\phi}} \hat{\bl{\phi}}$
is the gravitational acceleration in the disk plane (including that of the disk and the star) and 
$\bl{\nabla}_{\mathrm{p}}=\hat{\bl{r}}\partial/\partial r+\hat{\bl{\phi}}r^{-1}\partial/\partial\phi$
is the gradient along the planar coordinates of the disk. 
Turbulent viscosity enters the basic equations via the viscous stress
tensor $\bl{\Pi}$. The total cooling/heating rate is expressed as
\begin{equation}
    Q_{\rm tot} = \left(Q_{\rm cont}+Q_{\rm H2}+Q_{\rm HD} + Q_{\rm chem} +Q_{\rm metal}\right) 2 H,
    \label{coolQ}
\end{equation}
where $H$ is the vertical scale height calculated assuming a local hydrostatic equilibrium in the gravitational field of the star and disk \citep[see][]{2009VorobyovBasu}, $Q_{\mathrm{cont}}$ is the rate of radiative continuum cooling (or heating) of gas and dust (including collisional transfer of energy between gas and dust), $Q_{\mathrm{H_{2}}}$ is the H$_{2}$ line cooling rate, $Q_{\mathrm{HD}}$ is the HD line cooling rate, $Q_{\mathrm{chem}}$ is the chemical cooling/heating rate (through chemical reactions), and $Q_{\mathrm{metal}}$ is the metal line cooling rate.  All constituents of $Q_{\rm tot}$ are volumetric cooling or heating rates. The dust temperature is determined in the steady-state limit by the energy balance on dust grains due to the thermal emission, absorption, and collision with gas.  We emphasize that cooling/heating described by Equation~(\ref{coolQ}) does not include the adiabatic cooling/heating and viscous heating, the latter two mechanisms are taken into account by the first and third terms on the right-hand side of Equation~(\ref{eq:energy3}). 

In addition to hydrodynamic equations~(\ref{eq:mass})--(\ref{eq:energy3}),
we solve non-equilibrium kinetic equations for H, H$_{2}$, H$^{+}$, D, HD, D$^{+}$, and e, while the H$^{-}$ fraction is calculated from the equilibrium assumption. We assume that helium is always neutral and its fractional abundance is $y_{\mathrm{He}}=8.333\times10^{-2}$. We further assume that our species are collisionally coupled  with gas and solve  the continuity equation for each considered species using 
the same third-order-accurate algorithm as for the gas.
A more detailed description of the thermal evolution scheme and the solution procedure can be found in \citet{2020VorobyovMatsukoba}.

The central protostar is not only a source of gravity, but also provides radiative heating through irradiation of the disk surface. 
Stellar characteristics, such as the radius and photospheric luminosity, are calculated in line with the disk evolution using the stellar evolution code STELLAR \citep{2008YorkeBodenheimer}, modified by \citet{2013HosokawaYorke} to take different metallicities into account. The stellar characteristics are further used to calculate the total (stellar photospheric plus accretion)  luminosity and the radiation flux impinging on the surface of the disk. The co-evolution of the disk and the central protostar allows us to accurately follow the evolution of  a highly dynamical system with variable accretion and luminosity.

The starting point of numerical simulations is the gravitational collapse of a prestellar core. We consider three model realizations with distinct metallicities: $Z=0.01~Z_\odot$, $Z=0.1~Z_\odot$, and $Z=1.0~Z_\odot$. In addition, models with different $\alpha$-parameters describing the strength of turbulent viscosity in the disk are considered. The initial radial profile of the gas surface density $\Sigma$ and angular velocity $\Omega$ of the pre-stellar core has the
form typical of rotating cloud cores formed via slow expulsion of magnetic field due to ambipolar diffusion,  with  the  angular  momentum  remaining  constant during  axially-symmetric  core  compression  \citep{1997Basu}: 
\begin{equation}
\Sigma=\frac{r_{0}\Sigma_{0}}{\sqrt{r^{2}+r_{0}^{2}}}\label{eq:sigma}
\end{equation}
\begin{equation}
\Omega=2\Omega_{0}\left(\frac{r_{0}}{r}\right)^{2}\left[\sqrt{1+\left(\frac{r}{r_{0}}\right)^{2}}-1\right]\label{eq:omega}
\end{equation}
where $\Sigma_{0}$ and $\Omega_{0}$ are the gas surface density and the angular velocity at the center of the core. 
The radius of the central plateau is proportional to the thermal Jeans length and is defined as
$r_{0}=A c_{\rm{s}}/\sqrt{\pi G \rho_0}$, where $\rho_0$ is the central gas volume density, $c_{\mathrm{s}}=\sqrt{{\cal R} T_{g,0}/\mu}$ is the initial sound speed in the core, $\mu$ is the mean molecular weight, $T_{g,0}$ is the initial gas temperature, $\cal{R}$ is the universal gas constant, and $A$ is the amplitude of the initial positive density perturbation that drives the system out of equilibrium and trigger gravitational collapse.

\begin{table*}
\center
\caption{\label{table:1}Model parameters}
\begin{tabular}{ccccccccccc}
 & &  &  &  &  &  &  &   & &   \tabularnewline
\hline 
\hline 
Model & Metallicity & $M_{\mathrm{core}}$ & $\beta$ &  $\Omega_{0}$ & $r_{0}$ & $\Sigma_0$ & $r_{\rm{out}}$ & $T_{\rm g,0}$ & $\alpha$ & $M_{\rm \ast,fin}$ \tabularnewline
 & $Z_\odot$ & [$M_{\odot}$]  & [\%] &  [$\mathrm{km\,s^{-1}\,pc^{-1}}$] & [au] & [$\mathrm{g\,cm^{-2}}$] & [pc] & [K] & & [$M_\odot$] \tabularnewline
\hline 
1\_Z1.0 & 1.0 & 0.88 & 0.5 &  2.2 & 1560 & 0.09 & 0.05 & 10.1 & $10^{-2}$ & 0.48
\tabularnewline
2\_Z1.0 & 1.0 & 0.5 & 0.48 &  1.8 & 1560 & 0.09 & 0.031 & 10.1 & $10^{-2}$ & 0.33
\tabularnewline
1\_Z0.1 & 0.1 & 1.66 & 0.43 &  2.9 & 1720 & 0.1 & 0.075 & 12.2 & $10^{-2}$ & 0.62
\tabularnewline
2\_Z0.1 & 0.1 & 1.05 & 0.5 &  2.2 & 1720 & 0.1 & 0.05 & 12.2 & $10^{-4}$ & 0.49
\tabularnewline
3\_Z0.1 & 0.1 & 1.05 & 0.5 &  2.2 & 1720 & 0.1 & 0.05 & 12.2 & $10^{-2}$ & 0.465
\tabularnewline
4\_Z0.1 & 0.1 & 0.5 & 0.43 &  1.2 & 1720 & 0.1 & 0.027 & 12.2 & $10^{-2}$ & 0.31
\tabularnewline
1\_Z0.01 & 0.01& 1.75 & 0.46 &  5.5 & 610 & 0.7 & 0.031 & 30.3 & $10^{-2}$ & 0.82
\tabularnewline
2\_Z0.01 & 0.01 &1.04 & 0.5 &  3.8 & 610 & 0.7 & 0.019 & 30.3 & $10^{-4}$ & 0.51
\tabularnewline
3\_Z0.01 & 0.01 & 1.04 & 0.5 &  3.8 & 610 & 0.7 & 0.019 & 30.3 & $10^{-2}$ & 0.63
\tabularnewline
4\_Z0.01 & 0.01 & 0.52 & 0.5 &  2.3 & 610 & 0.7 & 0.011 & 30.3 & $10^{-2}$ & 0.39
\tabularnewline

\hline 
\end{tabular}
\center{ \textbf{Notes.} $M_{\mathrm{core}}$ is the initial core
mass, $\beta$ is the ratio of rotational to gravitational energy of the core,  $\Omega_{0}$ and $\Sigma_0$ are the angular velocity and gas surface density at the center of the core, respectively, $r_{0}$ is the radius
of the central plateau in the initial core, $r_{\mathrm{out}}$ is the initial radius of the
core, $T_{\rm g,0}$ is the initial gas temperature, $\alpha$ is the value of the viscous parameter, and $M_{\rm \ast,fin}$ is the final stellar mass at the end of computations.}
\end{table*}

To define the initial surface densities, temperatures, and chemical fractions of prestellar cores at distinct metallicities, we use a thermal model that is similar to that presented by \citet{2005Omukai} for collapsing cores in the one-zone approximation. Their figure~1 presents the gas temperature ($T_{\rm g}$) evolution of prestellar clouds with different metallicities as a function of the gas number density $n$ in the center of the core. For $Z=1.0~Z_\odot$ and $Z=0.1~Z_\odot$ we choose $n=10^6$~cm$^{-3}$, while for $Z=0.01~Z_\odot$ we use $n=2\times 10^7$~cm$^{-3}$. A higher value of $n$ for the latter case is explained by the fact that at lower densities the $T_{\rm g} \propto n^\gamma$ relation is steeper (i.e., $\gamma$ is higher) than what is allowed for gravitational collapse to occur ($\gamma\le 4/3$). At lower densities, the $Z=0.01~Z_\odot$ core would not collapse without strong external compression, which is not considered in our model.

Once the central number density is set, the corresponding gas and dust temperatures are obtained, and the resulting $r_0$ and $\Sigma_0=\mu m_{\rm H} n r_0$ are used to define the radial surface density profile of the core, where $m_{\rm H}$ is the mass of hydrogen. We note that \citet{2005Omukai} neglected the external irradiation heating other than that provided by the cosmic microwave
background. Here, we also take external stellar irradiation into account with a constant temperature of 10~K. This explains why our solar-metallicity core has a higher temperature than what was obtained in \citet{2005Omukai}.
The outer radius of the core is varied to create objects of different mass. The initial angular velocity profile is set by varying $\Omega_0$ until the desired ratio of rotational to gravitational energy $\beta$ is obtained. 

The initial parameters of the models are detailed in Table~\ref{table:1}. We consider two models with the solar metallicity and eight additional models with lower metallicities. The models are dubbed according to their ordinal number and metallicity. For instance, model~2\_Z0.1 corresponds to model 2 with metallicity $Z=0.1~Z_\odot$. The initial core masses of models with low metallicity vary by a factor of three to produce disks with different masses.  The ratio of rotational-to-gravitational energy $\beta$ is similar in all models to exclude its effect on the disk evolution. At the same time, $\beta$ is sufficient to produce gravitationally unstable disks in the solar metallicity case \citep{2013Vorobyov}. 

The numerical resolution is $512\times 512$ grid zones for all models except for models~4\_Z0.1 and 4\_Z0.01 with smallest prestellar cores, for which a twice coarser resolution was used.  The polar grid ($r,\phi$) is logarithmically spaced in the radial direction and has equal spacing in the azimuthal direction. The inner 5~au were cut out and replaced with the sink cell.
 In this work, the mass accretion 
rates  are calculated as the gas mass that crosses the sink--disk interface per unit time. We note that the actual mass accretion rate on the star may be modified by the physical
conditions and mechanisms that may operate in the innermost disk regions (such as the magnetorotational instability). The radial size of the innermost grid cell (just outside 5~au) varies in the 0.075--0.14~au limits, depending on the model, and it scales linearly with the radial distance.  The Truelove criterion states that the local Jeans length must be resolved by at least four numerical cells to correctly capture disk fragmentation \citep{1998Truelove}. In the thin-disk limit the Jeans length can be expressed as \citep{2013Vorobyov}
\begin{equation}
R_{\rm J} = {c_{\rm s}^2 \over \pi G \Sigma}.
\end{equation}
Fragments usually condense out of the densest sections of spiral arms at a typical distance of 100~au and then either migrate inward or scatter outward. The typical surface densities and temperatures in spiral arms do not exceed 100~g~cm$^{-2}$ and 100~K.
Adopting these values, the corresponding Jeans length is 
$R_{\rm J}\approx$ 20~au. The numerical resolution at 100~au varies in the 1.5--2.8~au limits, thus fulfilling the Truelove criterion. Finally, we note that the inner boundary condition allows matter to flow in both directions through the sink-disk interface, thus greatly reducing a spurious drop in the gas density near the sink that may develop in models with a one-way boundary condition  \citep[for details, see][]{2018VorobyovAkimkin}.

\begin{figure}
\begin{centering}
\includegraphics[width=1\columnwidth]{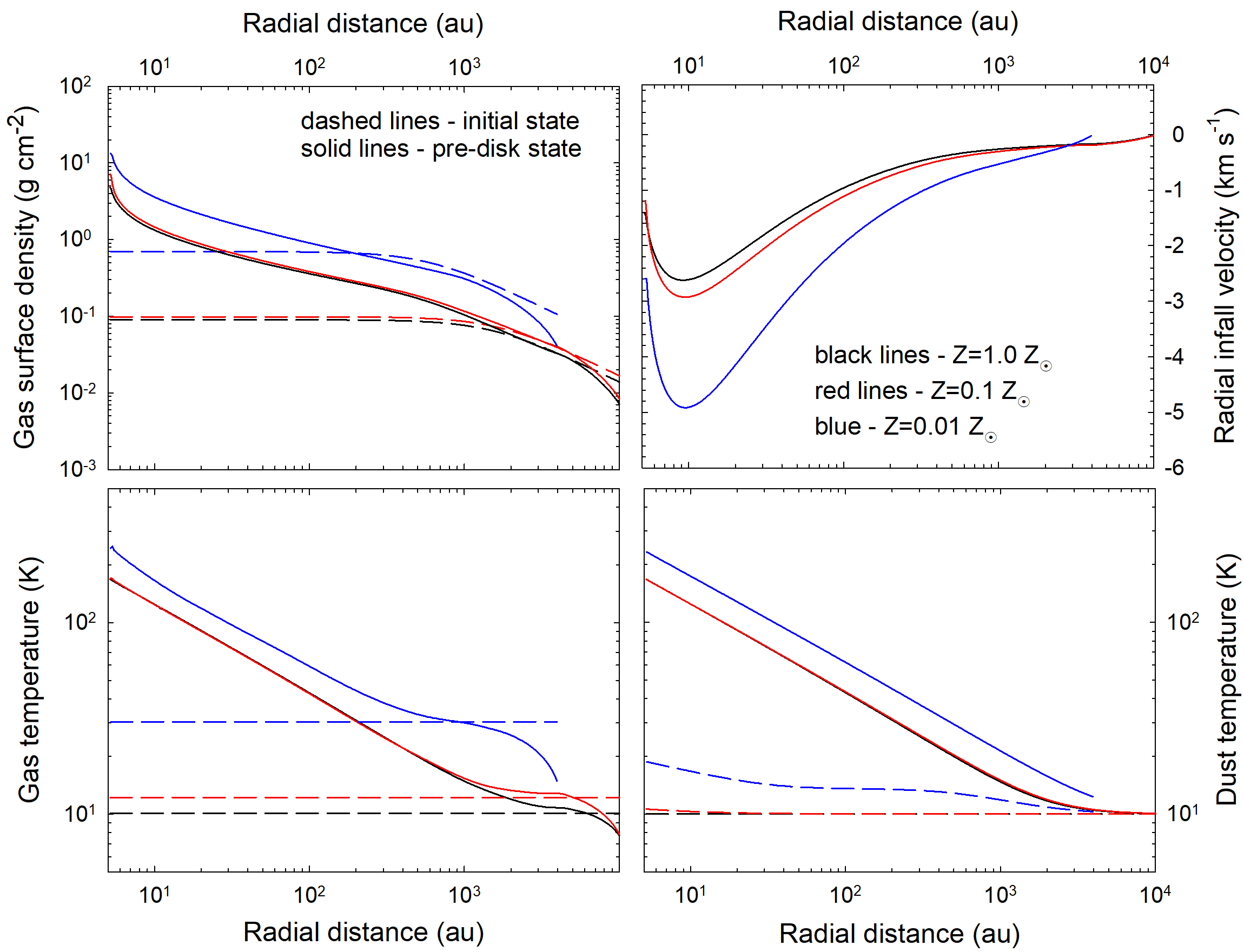}
\par\end{centering}
\caption{\label{fig:0} Radial profiles of the gas surface density (top-left), radial infall velocity (top-right), gas temperature (bottom-left), and dust temperature (bottom-right) in model~Z1.0. The dashed curves correspond to the initial setup at the onset of gravitational collapse, while the solid curves show the quantities at the time instance immediately preceding disk formation. Curves of different color correspond to models with different metallicities as shown in the legend.}
\end{figure}

\section{Pre-disk stage}
\label{predisk}
In this section, we focus on the initial stages of core collapse for different metallicities. Figure~\ref{fig:0} depicts the radial distributions of gas surface density, gas and dust temperatures, and radial infall velocity at the onset of collapse and at the time instance immediately preceding disk formation. Three models with similar core masses but distinct metallicities are shown: model~1\_Z1.0 (black lines), model~2\_Z0.1 (red lines) and model~2\_Z0.01 (blue lines). The dashed lines correspond to the initial state and the solid lines shows the pre-disk state. 

There are several key differences in the prestellar evolution of the $Z=0.1-1.0~Z_\odot$ models, on the one hand, and $Z=0.01~Z_\odot$ model, on the other hand. First, the gas and dust temperatures in the latter model are systematically higher than in the former models. At the considered metallicities and densities, the dust continuum emission is a dominant cooling mechanism \citep{2005Omukai} and the reduction in the dust content leads to an increase in the gas temperature during the collapse phase. 

Furthermore, the dust temperature in model~2\_Z0.01 decouples notably from that of gas. In particular, in the initial state the decoupling is seen throughout the entire extent of the core, while in the pre-disk phase the decoupling is limited to the outer low-density regions. The effect is also noticeable at higher metallicities, but is strongest for $Z=0.01~Z_\odot$. Assuming  that the dust radiative cooling balances heating due to collisions with gas, the dust temperature can be expressed as \citep{2020VorobyovMatsukoba}
\begin{equation}
    T_{\rm d}\simeq 120~\mathrm{K} \left( {T_{\rm g} \over 100~\mathrm{K}}~\right)^{0.3} \left( {n \over 10^{10}~\mathrm{cm}^{-3}} \right)^{0.2}.
\end{equation}
By setting $T_{\rm g} = T_{\rm d}$ we define the threshold temperature above which gas and dust thermally decouple from each other. This threshold temperature can be written as 
\begin{equation}
\label{Tcrit}
T_{\rm crit} \simeq 130~\mathrm{K} \left( {n \over 10^{10}~\mathrm{cm}^{-3}} \right)^{0.3}. 
\end{equation}
Consider now model~2\_Z0.01 with $n=2\times 10^7$~cm$^{-3}$ in the initial state. The corresponding $T_{\rm g}=30.3$~K and Equation~(\ref{Tcrit}) yields $T_{\rm crit}=20$~K, meaning that the gas and dust temperatures decouple from each other. This is indeed the case,  as is seen in Figure~\ref{fig:0}.  For model~1\_Z1.0 with 
$n=10^6$~cm$^{-3}$ in the initial state the critical temperature is $T_{\rm crit}=8.2$~K and the gas temperature (set by external stellar irradiation) is 10~K, meaning that decoupling is weak. In the pre-disk state the trend is similar -- lower metallicity models show stronger decoupling between gas and dust temperatures.

Finally, we note that the infall velocity $v_{\rm r}$ in model~2\_Z0.01 is appreciably higher (by the absolute value) than in the other two models. This is in agreement with analytical arguments stating that 
the mass infall rate is related to the gas temperature as \citep{1977Shu}
\begin{equation}
    \dot{M}_{\rm infall}=-2\pi r \Sigma v_{\rm r} \simeq {T_{\rm g}^{3/2} \over G}.
    \label{Minfall}
\end{equation}
As we will see later in the text, higher infall rates and temperature decoupling  will have important consequences for the disk evolution in the lowest metallicity models.

\begin{figure}
\begin{centering}
\includegraphics[width=1\columnwidth]{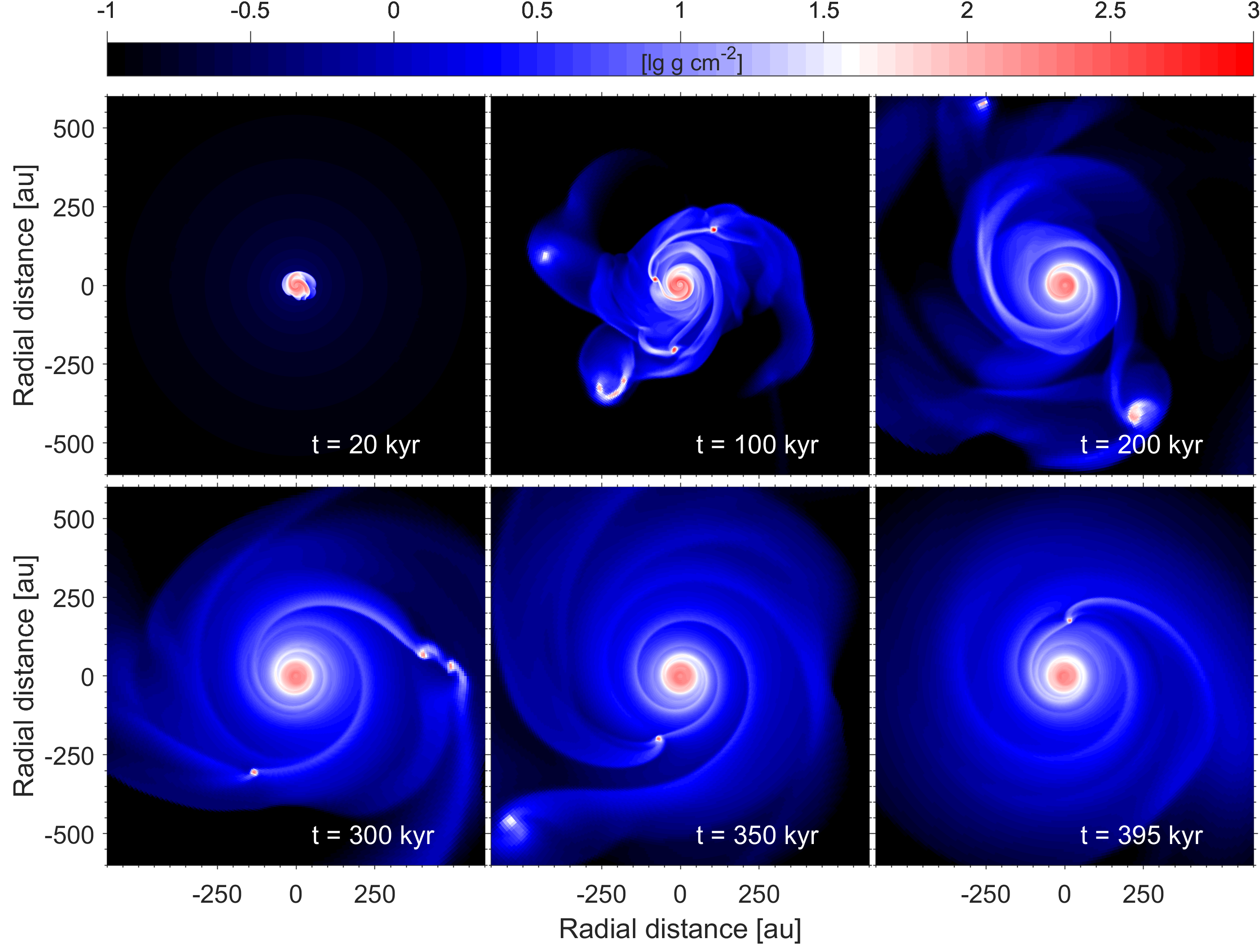}
\par\end{centering}
\caption{\label{fig:1} Disk evolution in model~1\_Z1.0. Shown is the gas surface density at six evolutionary times since the disk formation instance. {\bf The time is counted starting from the instance of disk formation, which occurs at 89~kyr in absolute time elapsed since the onset of gravitational collapse.} The scale bar is in g~cm$^{-2}$.}
\end{figure}

\section{Disk evolution and accretion bursts at solar metallicity}
\label{SolarZ}
In this section we review the disk evolution and protostellar accretion in the solar-metallicity case using model~1\_Z1.0 as a prototype example. This will help us later to understand the corresponding processes in low-metallicity models. Figure~\ref{fig:1} presents the gas surface density distribution in the inner $1200\times 1200$~au$^2$ box at six time instances after disk formation. Already after 20~kyr the disk becomes gravitationally unstable as indicated by the presence of two spiral arms. At this stage, the disk is nevertheless too small to fragment. Eighty thousand years later the disk is vigorously unstable and hosts several dense gaseous clumps formed via gravitational fragmentation. Disk fragmentation continues to the end of our numerical simulations for at least 400~kyr. We note that the number of fragments changes with time -- the increase means disk fragmentation and decrease implies clump destruction.  The properties of the formed clumps will be considered in a follow-up study. 

\begin{figure}
\begin{centering}
\includegraphics[width=1\columnwidth]{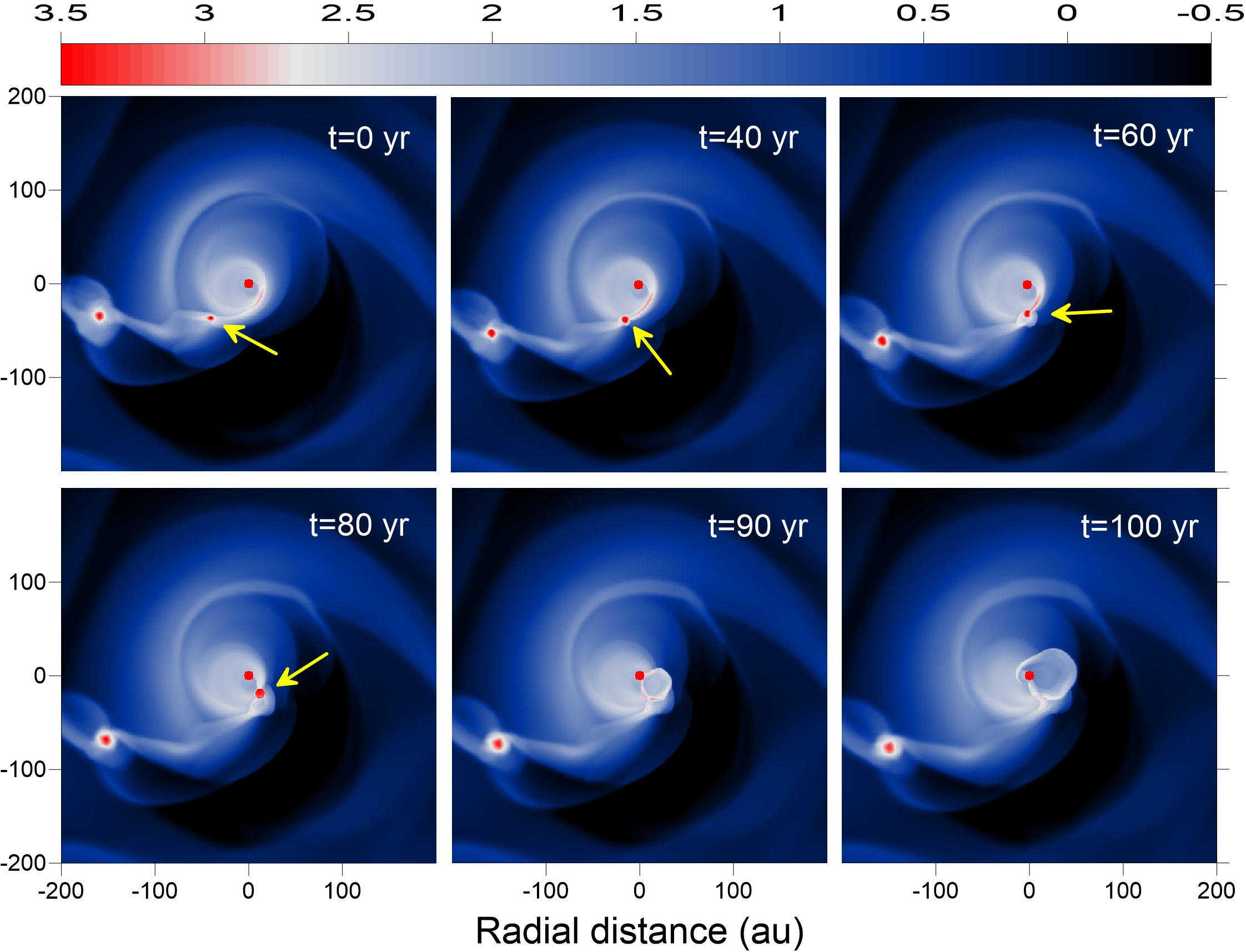}
\par\end{centering}
\caption{\label{fig:2} Disk evolution in model~Z1.0. Shown is the gas surface density at six evolutionary times encompassing a short time period of clump migration at $t\approx$~240~kyr.  The scale bar is in g~cm$^{-2}$.}
\end{figure}

Here, we focus on the inward migration of the clumps and on the accretion bursts that follow as the clumps disperse in the innermost disk regions. Figure~\ref{fig:2}
presents the disk evolution over a short period of time focused on one of such episodes of clump migration at $t\approx 240$~kyr. Shown is the inner $400\times 400$~au$^2$ box at six times spanning a range of 100~yr. The inward migrating clump is highlighted with the arrows. Initially, the clump is found at a radial distance of 54~au and after 80~yr it moved to 25~au. The quick inward migration is caused by the gravitational interaction between the clumps, as was described in detail in \citet{2018VorobyovElbakyan}. As the clump continues its migration towards the star its Hill radius becomes smaller than its radius and the clump starts losing its material. The Hill radius is defined as
\begin{equation}
    R_{\rm H} = r_{\rm cl} \left( {1\over 3} {M_{\rm cl} \over M_\ast+M_{\rm cl} } \right)^{1/3},
\end{equation}
where $M_{\rm cl}$ is the clump mass, $M_\ast$ is the stellar mass, and $r_{\rm cl}$ is the radial position of the clump.
Indeed, at t=80~yr the Hill radius is 4.5~au, while the radius of the clump is approximately 8~au, meaning that the clump has already begun to disintegrate. The hollow structure at about the position expected for the clump at 90 and 100 yrs is related to this burst-like disintegration, perhaps caused by tidal heating of the fast-rotating clump. Part of the released material is accreted through the central sink cell causing an accretion burst. The process of clump migration  repeats as long as the disk is massive enough to sustain continual disk fragmentation. This usually occurs in the embedded stage of disk evolution (lasting for several hundred kyr) when mass-loading from the infalling envelope replenishes the disk mass loss via accretion.

\begin{figure}
\begin{centering}
\includegraphics[width=1\columnwidth]{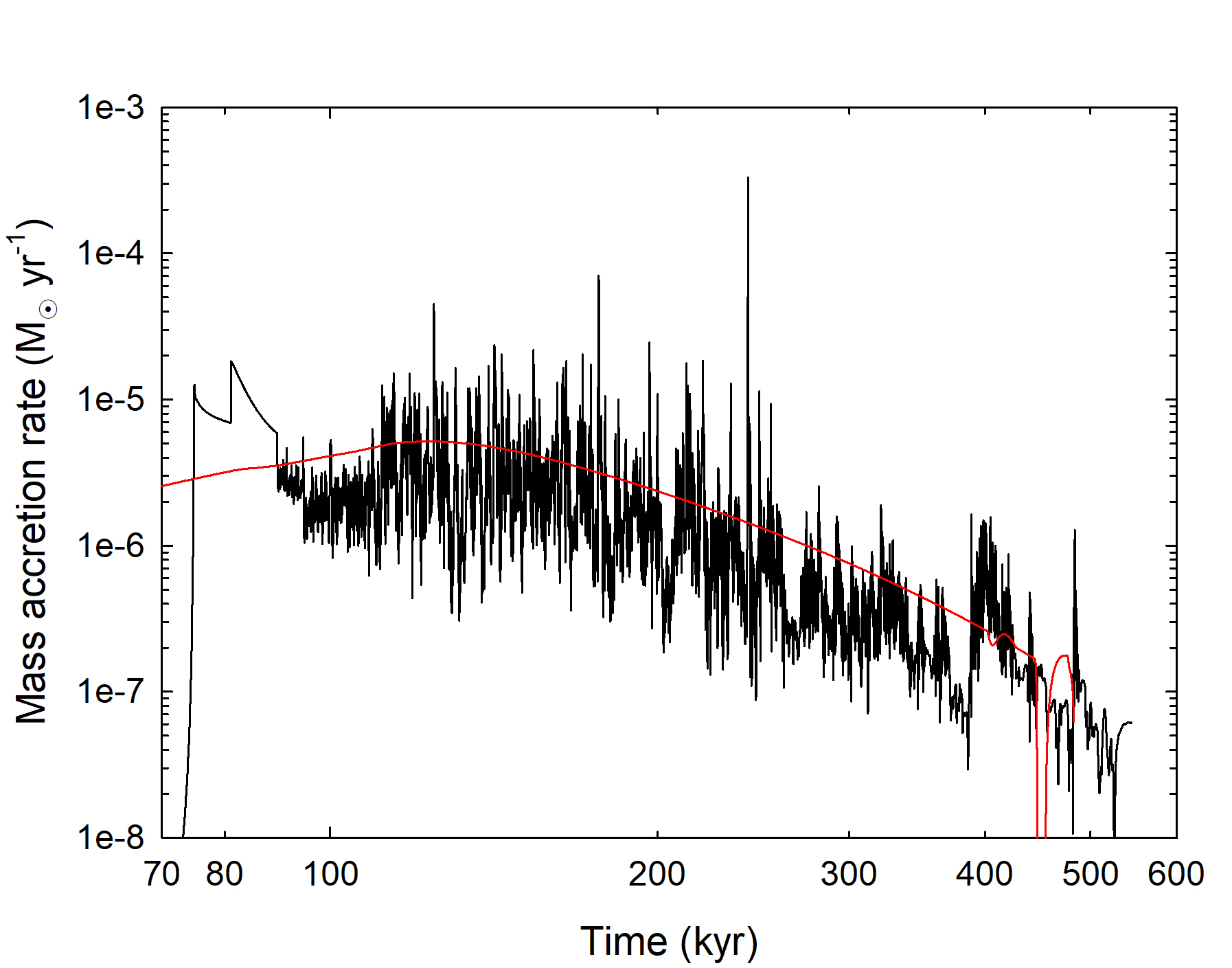}
\par\end{centering}
\caption{\label{fig:3} Mass accretion rate vs. time in model~1\_Z1.0 shown with the black line. The red line presents the mass infall rate on the disk. The time is counted from the onset of gravitational collapse and is shown on the log scale to better resolve the variability at early times. The disk forms at 89~kyr. }
\end{figure}

The black line in Figure~\ref{fig:3} presents the mass accretion rate as a function of time in model~1\_Z1.0, which is calculated as the mass passing through the sink cell per unit time.  We emphasize that the radius of the sink (5~au) is much greater than the stellar radius,  meaning that the calculated accretion rates may be affected by the disk physics interior to 5~au. For instance, the magnetorotational instability may introduce additional accretion bursts \citep{2009ZhuHartmannGammie} or the infalling clumps themselves may act as a trigger for the magnetorotational instability 
\citep{2014Ohtani}. 
The red line shows the mass infall rate on the disk from the envelope. The latter quantity is calculated as the mass flux at a radial distance of 1000~au from the star. The mass accretion rate is both time-declining and highly variable. High accretion variability reflects a highly dynamical nature of disk evolution. Long-term variations are caused by secular changes in the spiral pattern \citep{2016Elbakyan}, while the high-magnitude short-term spikes are triggered by infalling clumps. The strongest burst at $\approx 240$~kyr corresponds to the clump infall depicted in Figure~\ref{fig:2}. The strength of accretion variability diminishes as the mass infall rate on the disk declines with time. Reduced disk mass-loading cannot anymore sustain gravitational instability and fragmentation in the disk, which is the main driving force for accretion variability in our model.  The reader is referred to \citet{2015VorobyovBasu} for the in-depth discussion of the accretion bursts caused by inward-migrating clumps in the gravitationally unstable solar-metallicity disks.  It is also worth mentioning that some clumps may avoid inward migration and tidal destruction, thus forming giant planets or brown dwarfs on orbits from tens to hundreds of astronomical units \citep[e.g.,][]{2011ChaNayakshin,2013Vorobyov,2015Stamatellos,2018VorobyovElbakyan}. This is consistent with the detection of giant planets and brown dwarfs on wide orbits via direct imaging \citep[e.g.,][]{2008MaroisMacintosh,2010LafreniJayawardhana,2017Chauvin}.

\section{Disk evolution and accretion burst at lower metallicities}
\label{LowZ}
\begin{figure}
\begin{centering}
\includegraphics[width=1\columnwidth]{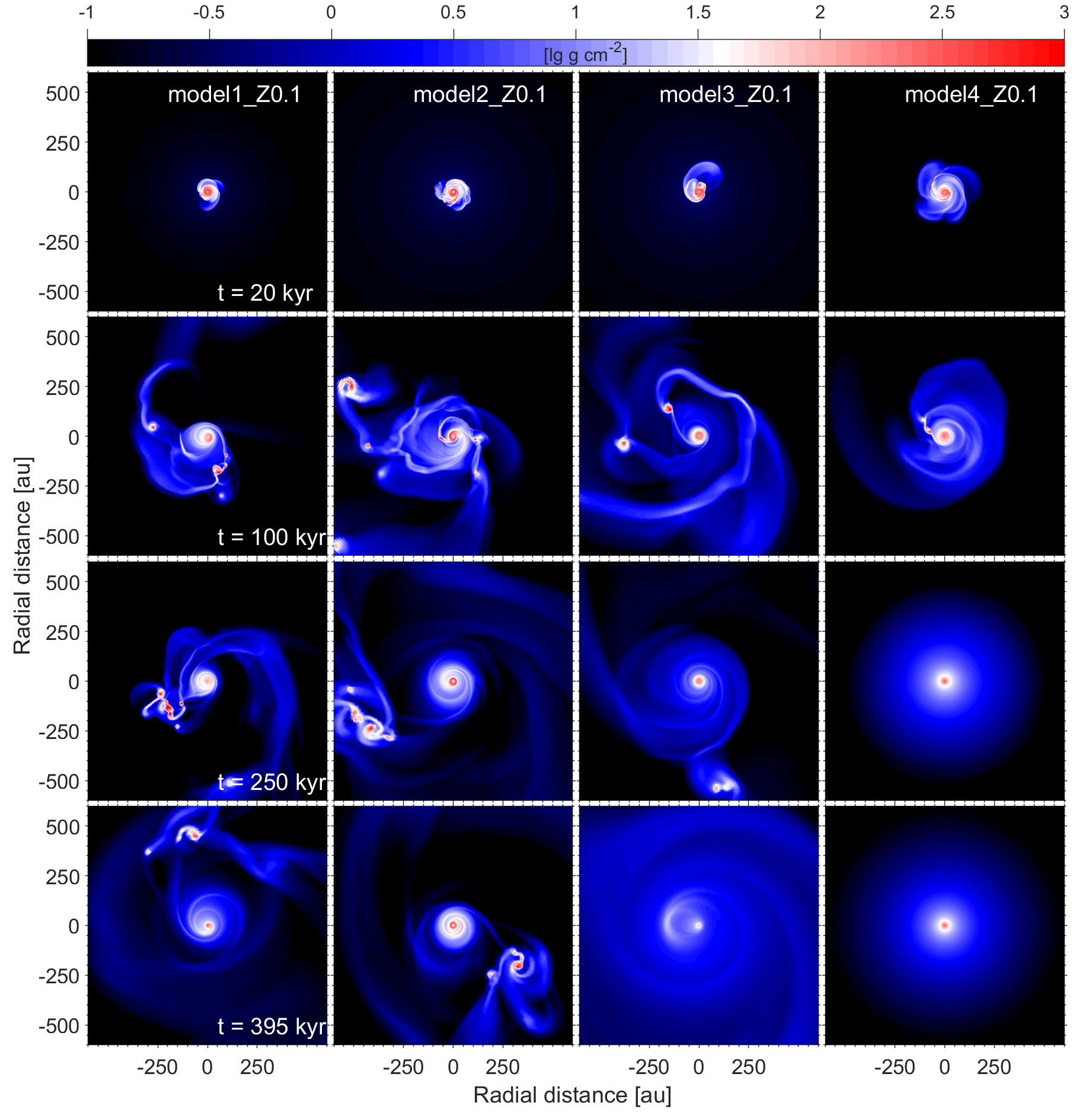}
\par\end{centering}
\caption{\label{fig:4} Disk evolution in models with metallicity ten times smaller than solar. Each column shows a particular model as indicated in the top row. Each row corresponds to a specific time since the disk formation instance as indicated in the left column.  }
\end{figure}

We now proceed with discussing the disk evolution in models with metallicities lower than solar. We start with the $Z=0.1~Z_\odot$ case and show in Figure~\ref{fig:4} the gas surface densities in the inner $1200\times 1200$~au$^2$ box for  the initial 395~kyr of disk evolution. All four models are presented, starting from the most massive model~1\_Z0.1 (left column, $M_{\rm core}=1.66~M_\odot$) and ending with the least massive model~4\_Z0.1 (right column, $M_{\rm core}=0.5~M_\odot$). Models~1\_Z0.1 and 2\_Z0.1 feature vigorous gravitational instability and fragmentation throughout the entire considered period of evolution. The disks in these models have fragmented into  highly dynamical clusters that comprise one massive and several less massive clumps. What remained of the original disk is now concentrated around the central star and is characterized by a notably smaller radius. The masses of the most massive clumps are 27.6 and 18.7~$M_{\rm Jup}$, indicating the possible formation of low-mass brown dwarfs on wide orbits.

On the other hand, models~3\_Z0.1 and 4\_Z0.1 show disk fragmentation only during the initial 100--200~kyr of evolution. Afterwards the disk becomes either weakly unstable or even nearly axisymmetric. This difference in the disk evolution between models 1\_Z0.1 and 2\_Z0.1, on the one hand, and models~3\_Z0.1 and 4\_Z0.1, on the other hand, can be explained by two factors. First, more massive cores produce more massive disks and this favours stronger gravitational instability and fragmentation in models 1\_Z0.1 and 2\_Z0.1. Second, turbulent viscosity represents an additional mechanism for mass transport, thus effectively reducing the disk mass via accretion  and spreading the disk outwards. Turbulent viscosity also represents an additional source of heating (the last term on the right-hand side of Equation~\ref{eq:energy3}), thus warming up the disk and reducing the strength of gravitational instability. These factors favour stronger instability in model~2\_Z0.1 with $\alpha=10^{-4}$ in comparison to the otherwise identical model~3\_Z0.1 but with $\alpha=10^{-2}$. A similar dependence of GI on the strength of turbulent viscosity was also reported by \citet{2018Rice}. 

When compared to the solar-metallicity case, no quantitative difference is seen in the disk evolution of $Z=0.1~Z_\odot$ models.
Disks with both metallicities develop gravitational instability and fragmentation, and the strength and longevity of these phenomena are similar when models with similar initial core masses are considered (e.g., model~1\_Z1.0 and model~3\_Z0.1). This is in agreement with what was found by \citet{2002Boss} and \citet{2010MeruBate}, but is in stark disagreement with the findings of \citet{2006Cai} who reported no disk fragmentation for sub-solar-metallicity disks. We note, however, that the latter authors did not consider mass-loading from the infalling envelope, which can qualitatively change the disk evolution and drive it to the fragmenting state \citep{2005VorobyovBasu,2008Kratter}. It appears that reducing the metallicity by a factor of 10 does not notably change the overall disk evolution, as was also the case for the pre-disk collapse phase in Figure~\ref{fig:1}.

\begin{figure}
\begin{centering}
\includegraphics[width=1\columnwidth]{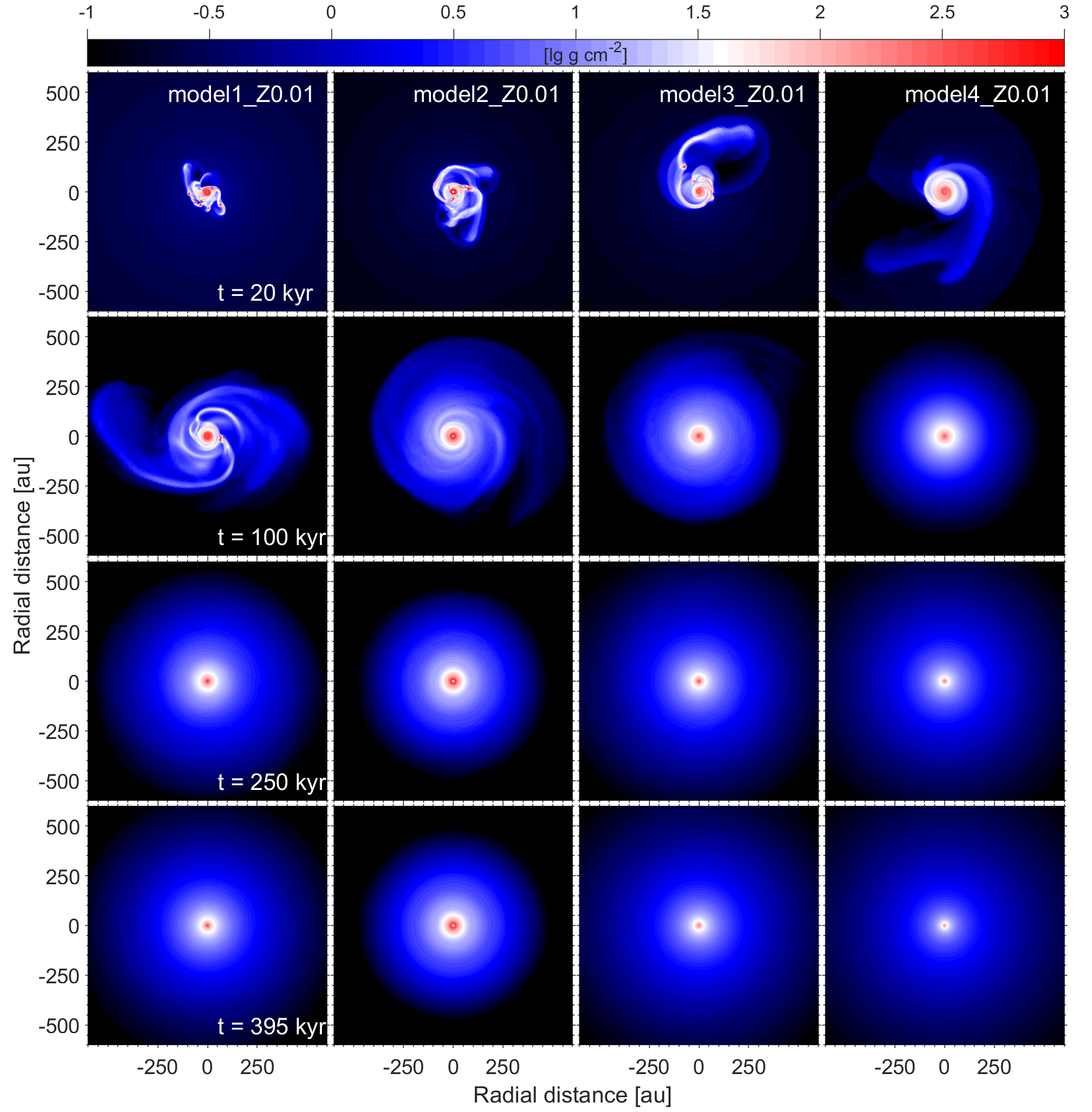}
\par\end{centering}
\caption{\label{fig:5} Similar to Figure~\ref{fig:4}, but for metallicity 100 times smaller than solar.  }
\end{figure}

Now we turn to the least metal-rich models of our sample.
Figure~\ref{fig:5} presents the disk evolution in four models with $Z=0.01~Z_\odot$ at the same time instances as for the models with $Z=0.1~Z_\odot$. We note that the initial masses and ratios of rotational to gravitational energy in both sets of models are similar. However, the disk appearance in the $Z=0.01~Z_\odot$ models is strikingly different. 
The disks in all models become virtually axisymmetric after 250~kyr of evolution. GI and fragmentation exist only in the very early stages of disk evolution. Model~4\_Z0.01 with the lowest initial core mass ($M_{\rm core}=0.52~M_\odot$) does not show disk fragmentation at all, while in the other models disk fragmentation ends by 100~kyr.  A notable exception is the most massive model~1\_Z0.01, for which disk fragmentation extends slightly beyond 100~kyr. The low-viscosity model with $\alpha=10^{-4}$ forms a notably more compact disk (no viscous spreading) and its disk appears to be slightly more unstable than its $\alpha=10^{-2}$ counterpart. 

The reason for the apparently different behavior between the $Z=1.0-0.1~Z_\odot$ disks and $Z=0.01~Z_\odot$ disks lies in the thermal evolution of the considered models. We showed in Figure~\ref{fig:0} that the models with $Z=0.01~Z_\odot$ have systematically higher gas temperatures, dust temperatures, and  infall velocities as compared to the higher-metallicity models. The immediate consequence is that the mass infall rate on the disk is higher by up to a factor of 5 in the lowest-metallicity models, affecting the entire disk evolution (note that the infall rate is proportional to the gas temperature to the power 3/2, see Eq.~\ref{Minfall}). Indeed, the comparison of Figures~\ref{fig:2} and \ref{fig:4}, on the one hand, and Figure~\ref{fig:5}, on the other hand, demonstrates the disk appears more developed at 20~kyr after its formation in models with the lowest metallicity. With a higher infall rate, the embedded phase in models with $Z=0.01~Z_\odot$ lasts much shorter (for the same mass reservoir in the prestellar core), and this also acts to shorten the gravitationally unstable phase. For instance, the embedded phase in model~3\_Z0.1 ends about 320~kyr after disk formation, while in model~3\_Z0.01 it ends about 40~kyr after disk formation. When making these calculations, we defined the end of the embedded phase as the time instance when less than 10\% of the initial core mass was left in the infalling envelope. We emphasize that a shorter embedded phase in the $Z=0.01~Z_\odot$ models does not prevent disk from fragmenting because disk fragmentation takes place on orbital timescales, which are still much shorter than the time extent of the embedded phase. After the embedded phase no accretion onto the disk takes place and the gas surface density cannot easily increase globally to sustain instability.

\begin{figure}
\begin{centering}
\includegraphics[width=1\columnwidth]{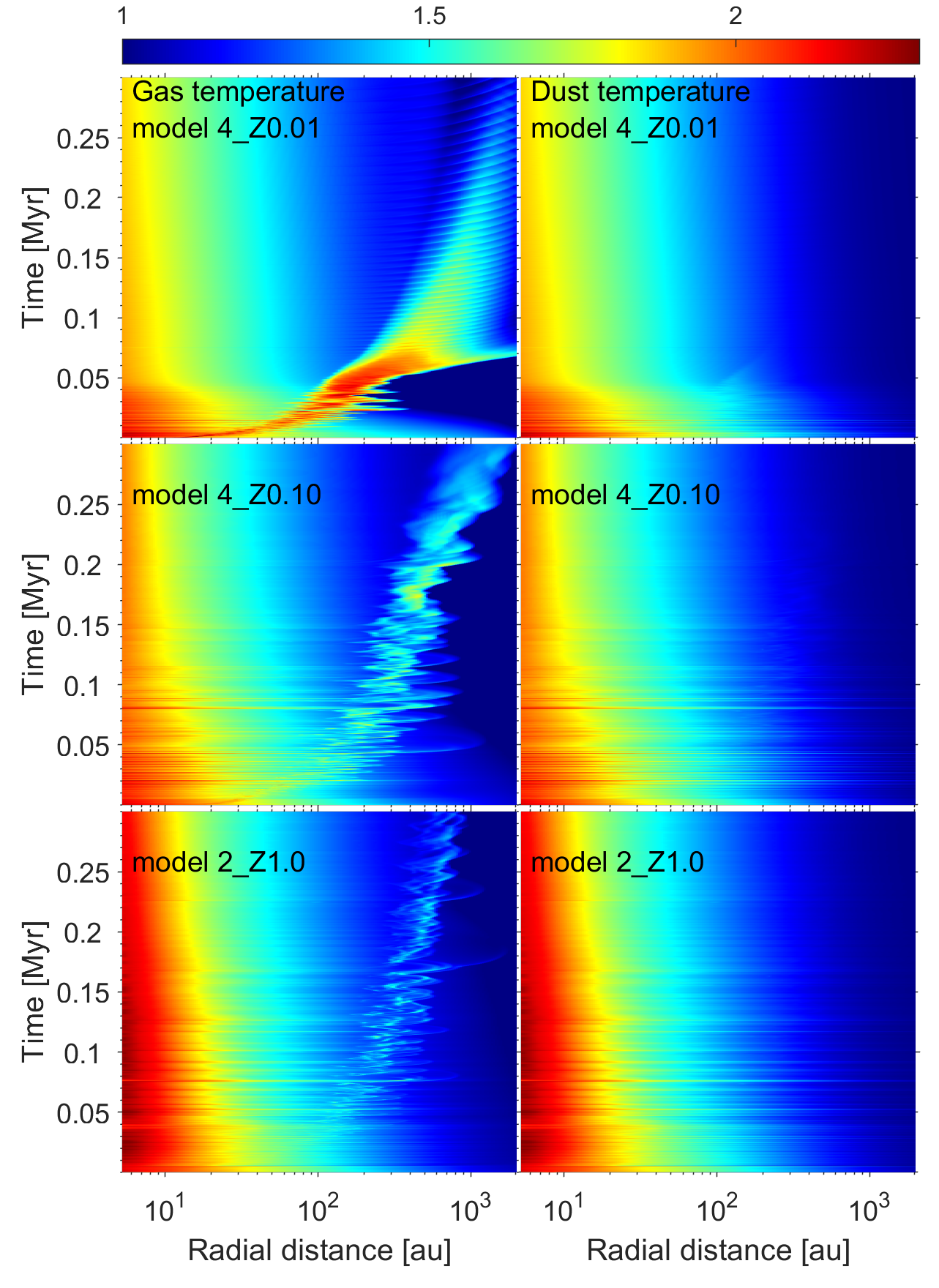}
\par\end{centering}
\caption{\label{fig:6} Spacetime diagrams of the gas temperature (left column) and dust temperature (right column) in models with different metallicities as indicated in each panel. The color bar is units of Kelvin on the log scale. The time is counted from the disk formation instance.}
\end{figure}

Moreover, the temperature decoupling between gas and dust can also contribute to the increased rate of gas infall on the disk in the lowest-metallicity models. As an example, we present in Figure~\ref{fig:6} the spacetime diagrams of the azimuthally averaged gas and dust temperatures in models 2\_Z1.0, 4\_Z0.1, and 4\_Z0.01. The initial 300~kyr of disk evolution is shown and the time is counted from the disk formation instance. These are the least massive models with disks having a regular (not highly fragmented) structure and are therefore best suited to demonstrate the temperature trends. We note that the horizontal high-temperature spikes are caused by accretion bursts discussed in detail in the next section.

The comparison between models with different metallicities  reveals two key differences in the gas and dust radial distribution. First, the higher metallicity disks are on average hotter in the inner several tens of astronomical units than their low-metallicity counterparts. This trend is caused by a lower optical depth  in low-metallicity disks. As a result, the viscous heat released in the disk midplane can more easily be radiated away by dust thermal emission. There are, however, subtle deviations from the described trend. Most notably, the $Z=0.01~Z_\odot$ model features a warmer disk in the initial stages of disk evolution ($t\le 0.05$~Myr) than the $Z=0.1~Z_\odot$ one. This is caused by a higher accretion rate and associated higher accretion luminosity (stellar irradiation) in the lowest metallicity model (see Fig.~\ref{fig:7} below). This high-accretion stage is, however, short-lived and is limited by the embedded phase. 

Second, the gas and dust temperatures decouple from each other in the vicinity of the disk outer edge and in the inner envelope. Moreover, the magnitude of temperature decoupling increases with decreasing metallicity and is remarkable in the $Z=0.01~Z_\odot$ case, as can be seen from a local peak in the gas temperature in the $10^2$--$10^3$~au region (such a peak is absent in the dust temperature distribution).  The temperature decoupling occurs in regions with low density (see Equation~\ref{Tcrit}) occupied by the outer disk and the inner envelope. The gas temperature there is notably higher than the dust temperature. As a result, the mass infall rate $M_{\rm infall}$ also increases  and this in turn makes the disk evolve faster for the same mass reservoir in the collapsing core.

\begin{figure}
\begin{centering}
\includegraphics[width=1\columnwidth]{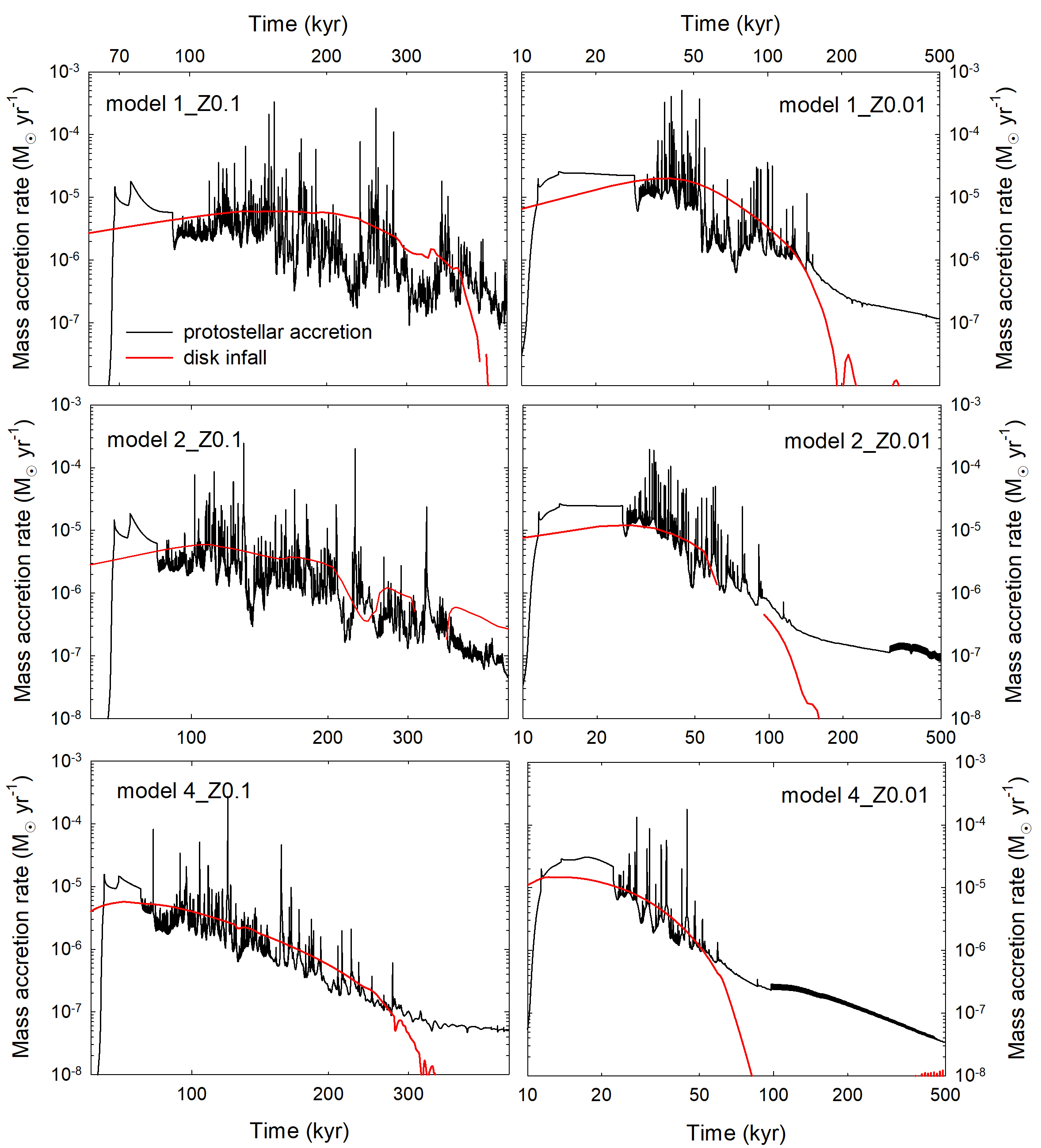}
\par\end{centering}
\caption{\label{fig:7} Mass accretion rate on the star (black line) and mass infall rates on the disk (red lines) vs. time in models with metallicities lower than solar. The left column presents models with metallicity $Z=0.1~Z_\odot$, while the right column does it for the $Z=0.01~Z_\odot$ metallicity. The time is counted from the onset of gravitational collapse and is shown on the log scale to better resolve the variability at early times. }
\end{figure}

In Sect.~\ref{SolarZ}, we demonstrated that gravitationally unstable disks of solar metallicity can be characterized by highly variable mass accretion rates with episodic bursts triggered by inward-migrating clumps. Here, we review the character of protostellar mass accretion in models with lower metallicities in the $Z=0.1-0.01~Z_\odot$ range. 
The black lines in Figure~\ref{fig:7} present the protostellar mass accretion rates ($\dot{M}$) as a function of time in models with $Z=0.1~Z_\odot$ (left column) and $Z=0.01~Z_\odot$ (right column).  The mass infall rates on the disk  ($\dot{M}_{\rm infall}$) are also plotted with the red lines. The latter quantity is calculated as a mass flux  at a radial distance of 1000~au from the star (all our model disks are smaller than 1000~au in radius). Several interesting features can be seen in the figure. 

Firstly, the mass accretion rate is highly variable in all models, but the variability period is limited  to the initial stages of disk evolution. The initial smooth behavior of $\dot{M}$ typical of the pre-disk phase is followed by a highly variable phase once the disk forms at several tens of kyr after the onset of gravitational collapse. The variability in the mass accretion rate diminishes together with a diminishing mass infall rate. As was noted in Sect.~\ref{SolarZ}, mass-loading from the envelope helps to replenish the disk material lost via protostellar accretion and keep the disk in the gravitationally unstable mode. Once the infall is over, the disk quickly attains a stable state that is characterized by a smoothly declining mass accretion rate. Secondly, the variability is stronger and the bursts are more numerous  in more massive models. This is again related to the strength of gravitational instability and fragmentation in more massive disks that form in models with initially more massive pre-stellar cores. Thirdly, the burst activity seems to be more pronounced in models with $Z=0.1~Z_\odot$ than in models with $Z=0.01~Z_\odot$. This is expressed in a longer time period of variability and a higher number of bursts.
Finally, both $\dot{M}$ and $\dot{M}_{\rm infall}$ are systematically higher in the $Z=0.01~Z_\odot$ models than in the $Z=0.1~Z_\odot$ ones during the embedded phase of disk evolution (40 and 320~kyr for $Z=0.01$ and $Z=0.1~Z_\odot$, respectively). A more rigorous statistical analysis of the burst, including their light curves and possible similarity to the classical FU-Orionis type bursts in the solar-metallicity environments, will be performed in a follow-up study.

\begin{figure}
\begin{centering}
\includegraphics[width=1\columnwidth]{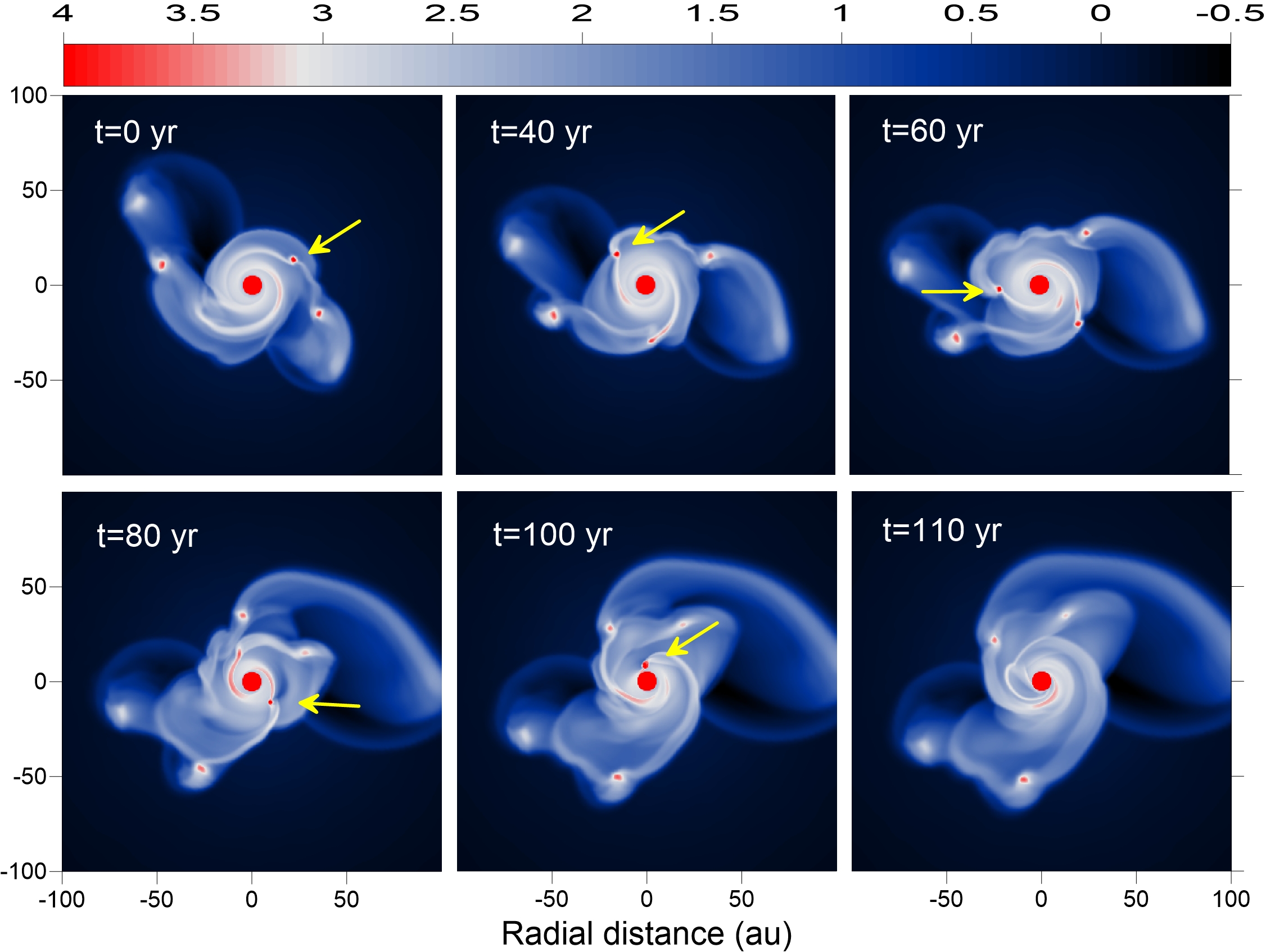}
\par\end{centering}
\caption{\label{fig:8} Inward migration and infall of a clump in the gravitationally unstable disk with $Z=0.01~Z_\odot$. Shown is the gas surface density in log g~cm$^{-2}$. The inward-migrating clump is highlighted by the yellow arrows. }
\end{figure}

Finally, we want to demonstrate the similarity between accretion burst mechanisms in different metallicity environments caused by disk fragmentation.
Figure~\ref{fig:8} provides an example of clump infall causing an accretion burst in model~1\_Z0.01. Shown is the gas surface density in the inner $200\times200$~au$^2$ box for a limited time period of 110~yr. The inward-migrating clump is highlighted by the yellow arrows. The overall picture is similar to what has been shown in Figure~\ref{fig:2} in the context of solar metallicity bursts. The clump spirals down towards the star until it is destroyed by tidal torques in the immediate vicinity of the sink cell. The mass of the clump at this stage is about $10~M_{\rm Jup}$, its radius is about 2~au, while the Hill radius is 1.5~au, meaning that the clump has started to lose its material, which is shortly accreted on the star producing a burst.   

A potential caveat needs to be mentioned with respect to this burst scenario. 
The numerical resolution of the clump is insufficient to resolve the second collapse (see Sect.~\ref{model}). This means that part of the clump may avoid tidal destruction if collapsed to protoplanetary densities and sizes \citep[see, e.g.,][]{2018VorobyovElbakyan}. In this case, the burst might be of a notably smaller magnitude than what was found here. These topics deserve detailed follow-up studies.

\section{Conclusions}
\label{summary}

We studied numerically the early stages of disk evolution and protostellar accretion in young stellar systems with metallicities from 10 to 100 times lower than solar. For this purpose we employed the numerical hydrodynamics simulations in the thin-disk limit that feature separate gas and dust temperatures and consider disk mass-loading from collapsing parental cloud cores. Low metallicities were set by scaling down the gas and dust opacities, dust-to-gas mass ratio, and metal content of the solar-metallicity disk by the corresponding factor.
To eliminate the possible uncertainty that may be caused by the initial conditions, we considered the gravitational collapse of prestellar cores with similar mass and ratio of rotational-to-gravitational energy, but distinct metallicity. A comparison with the solar-metallicity models is also performed. Our findings can be summarized as follows.

-- The initial stages of cloud core collapse are distinct in models within the considered metallicity range. In particular, the $Z=0.01~Z_\odot$ models are characterized by higher temperatures and infall velocities than their higher metallicity counterparts ($Z=1.0-0.1~Z_\odot$). The gas and dust temperatures decouple notably in the lowest metallicity model. 

-- The initial stages of disk evolution are characterized by vigorous gravitational instability and fragmentation. The time period that is covered by this unstable stage is, however, much shorter in the $Z=0.01~Z_\odot$ models as compared to their higher metallicity counterparts. This can be explained by elevated mass infall rates on the disk in models with the lowest metallicity, causing the corresponding disks to evolve faster for the same initial mass reservoir in the prestellar cores. Once the disk mass-loading from the infalling envelope diminishes (and this occurs also sooner in lower-metallicity models), the disk quickly acquires a gravitationally stable state.

-- Protostellar accretion rates are highly variable in the low-metallicity models reflecting the highly dynamical state of the corresponding protostellar disks. Similar to the solar-metallicity models, the $Z=0.1-0.01~Z_\odot$ systems feature short, but energetic episodes of mass accretion caused by infall of inward-migrating gaseous clumps that form in the outer disk via gravitational fragmentation. The bursts seem to be more numerous and lasting longer in the $Z=0.1~Z_\odot$ models in comparison to the $Z=0.01~Z_\odot$ case.

\section*{Acknowledgements}
We are thankful to the anonymous referee for useful comments that helped to improve the manuscript.
E.I.V. and M.G. acknowledge support from the Austrian Science Fund (FWF) under research grant 
P31635-N27. V.G.E. acknowledges the Swedish Institute for a visitor grant allowing to conduct research at Lund University. The simulations were performed on the Vienna Scientific Cluster.

\bibliographystyle{aa}
\bibliography{refs}

\end{document}